# Towards robust paralinguistic assessment for real-world mobile health (mHealth) monitoring: an initial study of reverberation effects on speech

*Judith Dineley[1], Ewan Carr[1], Faith Matcham[2], Johnny Downs[1], Richard Dobson[1,3],*
*Thomas F. Quatieri[4], Nicholas Cummins[1]*

[1]Institute of Psychiatry, Psychology and Neuroscience, King's College London, London, UK
[2]School of Psychology, University of Sussex, Falmer, UK
[3]Institute of Health Informatics, University College London, London, UK
[4]MIT Lincoln Laboratory, Lexington, MA, USA

`judith.dineley@kcl.ac.uk, nick.cummins@kcl.ac.uk`

## Abstract

Speech is promising as an objective, convenient tool to monitor health remotely over time using mobile devices. Numerous paralinguistic features have been demonstrated to contain salient information related to an individual's health. However, mobile device specification and acoustic environments vary widely, risking the reliability of the extracted features. In an initial step towards quantifying these effects, we report the variability of 13 exemplar paralinguistic features commonly reported in the speech-health literature and extracted from the speech of 42 healthy volunteers recorded consecutively in rooms with low and high reverberation with one budget and two higher-end smartphones, and a condenser microphone. Our results show reverberation has a clear effect on several features, in particular voice quality markers. They point to new research directions investigating how best to record and process in-the-wild speech for reliable longitudinal health state assessment.

## 1. Introduction

Speech processing applied to in-the-wild smartphone recordings in health state assessment has potential as a convenient, objective monitoring tool in health care and research [1]. As a clinical application, this may break down barriers affected individuals face in accessing healthcare. Multiple studies have demonstrated the inherent value of the paralinguistic content of speech in the detection and classification of a range of health disorders e.g., [2]–[8]

However, much of this research has used one-off speech recordings made in highly controlled environments with high-specification audio equipment, with collection overseen by a trained researcher. In-the-wild longitudinal monitoring with mobile devices, in contrast, introduces multiple technical, acoustic and human factors, and associated variability in the recorded speech signal, some of which could be erroneously interpreted as related to a change in health state. These factors may include the specification of the recording hardware, ambient noise and room acoustics. Smartphone design varies greatly, resulting in varying data quality and reproducibility issues when recording on different devices [9].

Literature quantifying sources of variability in recorded speech for mobile health applications is relatively sparse. Most studies have focused on dysphonia and a small number of features, most commonly comparing the performance of mobile devices in capturing speech features to a studio microphone [11]–[14]. Studies have shown that fundamental frequency (F0) is more robust to device type and ambient noise, while voice quality features such as jitter and shimmer are heavily impacted [12], [13]. Device type has also been shown to affect the signal-to-noise ratio of recorded speech [15]. Voice quality features have also been shown, in a controlled phonetics study, to be affected by recording environment [16].

The effects of ambient noise and microphone type have been somewhat investigated, however, the effect of reverberation on speech features in-the-wild has received no specific attention in the literature, to the best of the authors' knowledge, and differences in reverberation are often audible. To begin to address this knowledge gap, our pilot measures the effect of reverberation on exemplar speech features commonly used in health-driven analyses, recording healthy volunteers in rooms with low and high reverberation, simultaneously with several devices. We chose to focus on the healthy state to create baseline data to facilitate the interpretation of reverberation-affected data from clinical cohorts in future work. We assessed features extracted with standard open-source processing tools to observe the sensitivity of the entire data pipeline to reverberation and not only recording hardware.

Our first results, that are part of a larger body of work, are reported here. We report two analyses. First, we assessed the effect of reverberation on features extracted at a suprasegmental level from recordings, a common approach for baseline machine learning experiments. Second, removing the effects of linguistic variability, we report changes in features with reverberation at the vowel level. We analyzed recordings from a benchmark condenser microphone and three smartphones to observe the effects of reverberation in each. Our aim was to investigate the implications of room reverberation for mHealth tool robustness in research and clinical practice.

## 2. Methodology

With no public dataset of mobile device recordings in controlled conditions with varying reverberation available, we created a new speech dataset. To facilitate rapid recruitment of a balanced cohort in the short time frame available, it was decided to not make these recordings publicly available. Speech features and basic non-identifiable, anonymized participant data will be available upon completion of the project per the conditions of the participants' consent. The recording and analysis pipeline is summarized in Figure 1.

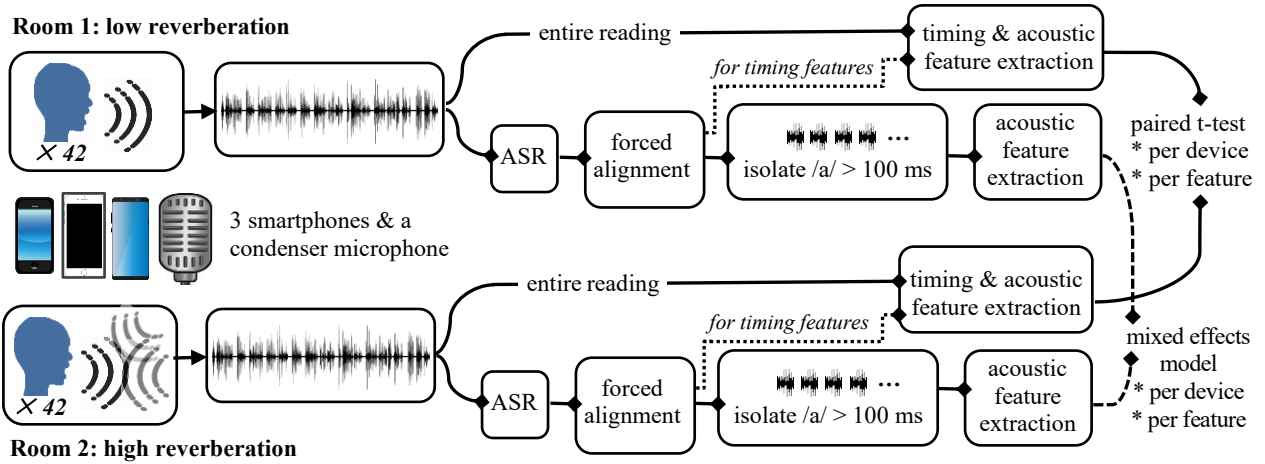

Figure 1: *Data pipeline for investigation of room reverberation effects on 13 exemplar timing and acoustic speech features extracted from smartphone recordings. Forced alignment provides information for timing feature extraction.*

### 2.1 Data Collection

**Participant recruitment:** This project received approval from the Research Ethics Committee of King's College London. Forty-two volunteers were recruited by word-of-mouth, social media and the institute's participant recruitment website and e-newsletter. Exclusion criteria included dyslexia, being a smoker, being under 16, ongoing treatment for a mental health disorder, and having any other kind of neurological, respiratory or other health issue that may affect the participant's speech. Volunteers whose first language was not English were required to have a sufficient level of reading ability and spoken English to read aloud two easy-intermediate texts and describe an everyday scene. Enrolment questionnaires recorded participants' age, sex, height, whether English was their first language and, if not, their level of English according to the Common European Framework of Reference for Languages[1]. For analysis, all participant data was saved and labelled only with a non-personally identifiable ID number.

**Speech Recording:** Participants were recorded in Summer 2022 in two small, neighboring, identical, windowless electroencephalogram test rooms with low ceilings in the basement of a university building where there was no ambient noise. *Room 1* was fitted with acoustic foam and soft furnishings and *Room 2* was empty, leaving only hard plaster and wooden surfaces, except for carpet tiles on the floor. Thus, we had recording spaces with low and high reverberation, respectively. In a single session, we recorded a participant consecutively in each room, alternating the room order between participants to avoid introducing bias between rooms. Participants read aloud the full version of *The Rainbow Passage* in each room [17] to elicit consistent speech across the cohort with minimal training. The reading provides variety and versatility for analysis, with high phonetic balance and full phonetic coverage, and structural and lexical complexity appropriate for our healthy, educated cohort [18], [19].

Each participant reading was recorded simultaneously with a Samsung Galaxy S20 FE 5G (released 2020), as a representative non-budget Android smartphone; a Motorola G6 Play (released 2018), as a representative budget Android smartphone; an Apple iPhone 11 (released 2019) and an Audio Technica AT2020USB+ condenser microphone with a cardioid pickup area, providing a higher quality recording from a non-mobile device within the budget constraints of our pilot study. The condenser microphone was fitted with a Rycote InVision shock mount and foam pop filter on a tabletop microphone stand and operated using Audacity open-source software (v. 3.1.3). The smartphones were positioned directly adjacent to the pop filter with their primary microphones at the height of the center of the condenser microphone diaphragm.

Participants were instructed to stand in front of and facing the pop filter. The height of the smartphones and condenser microphone was adjusted to be approximately level with the participant's mouth. Immediately prior to recording, the participant was positioned 30 cm from the condenser microphone, the distance at which its frequency response is specified. Upon completion of recording, audio files were saved and transferred to a secure cloud storage service. As with all participant data, recordings were labelled only with the participant's non-personally identifiable ID number.

### 2.2 Feature extraction and analysis

**Feature extraction:** We extracted 13 exemplar features commonly used in speech-health research that represent the basic speech production characteristics of timing, prosody, quality and articulation. These were speaking rate; articulation rate; number of pauses; mean fundamental frequency, (F0 mean); standard deviation (F0 standard deviation); intensity; first formant mean frequency, (F1 mean); second formant mean frequency, (F2 mean); spectral slope; spectral tilt; jitter; shimmer; and cepstral peak prominence, (CPP).

First, we converted all audio files to single channel 16 kHz Waveform Audio File Format (WAV) files. Audio files were then automatically transcribed, using the base Open AI Whisper model (v. 20230117)[2], choosing automated analysis over manual transcription as a more realistic solution for a real-world mHealth pipeline [20]. We aligned transcripts with the reading text with the Montreal Forced Aligner [21] and English MFA acoustic model V2.0.0a[3] to extract the three timing features for the entire reading.

---

[1] Council of Europe, www.coe.int/lang-cefr
[2] Radford et al, arxiv.org/abs/2212.04356
[3] mfa-models.readthedocs.io

Table 1: *Participant characteristics (n = 42)*

| sex | female | 23 | English | Yes | 29 |
|---|---|---|---|---|---|
|  | male | 19 | L1 | No | 13 |
| age | median | 28 | height | median | 1.70 |
| (years) | IQR | 23-32 | (m) | IQR | 1.63-1.80 |

Acoustic features were extracted at two levels: (1) suprasegmentally, at the level of the entire reading, and (2) at the vowel level, for occurrences of /a/ of at least 100 ms. We undertook a suprasegmental approach as this is a common approach in paralinguistic analyses [22] and enables the calculation of timing features. In contrast, analysis of /a/ vowel sounds removes variability due to the diverse linguistic content of the entire reading. An open vowel, /a/ has been recommended as more reliable for jitter and shimmer measurements [23], two features associated with changes in health, in particular mental state [7]. Timings provided by forced alignment enabled the extraction of feature values for each detected /a/ sound.

Acoustic features were extracted using Parselmouth (v. 0.4.1), an open-source Python library (v. 3.9.13), in combination with Praat (v. 6.3.02) [24], [25]. This extraction used default Praat settings, except for the extraction of F0, which followed the two-step approach recommended in [26]. The use of Whisper, Montreal Forced Aligner and Praat was a design choice to observe the sensitivity of feature extraction to reverberation effects using a processing pipeline of standard, open-source, well-established tools.

**Statistical analysis:** Tests were undertaken in R (v. 4.2.2). We analyzed the 13 features extracted at the suprasegmental level using bootstrapped paired t-tests. First, for each feature and device combination, we grouped the 84 feature values obtained for every participant in the two rooms, standardizing the values (mean = 0; standard deviation = 1). In each feature-device group, we then paired feature values per participant (Room 1, Room 2) to perform the t-tests and plotted standardized mean differences and corresponding 95% confidence intervals (CI).

We used linear mixed effects (LME) models [27] to analyze the effects of reverberation on the 10 acoustic features extracted from repeated /a/ vowel sounds identified in each of our recordings. We detected a median of 5 /a/ sounds ≥ 100 ms per recording (interquartile range: 3-10). Individual models were fitted for each feature-device combination, using groups of standardized feature values for each room-participant permutation, similar to the suprasegmental analysis. In contrast to the suprasegmental t-tests, LME models account for the clustering of repeated measures and allows us to adjust for participant gender, age, height and first language [28].

Each model considered a single speech feature as the dependent variable and included reverberation (0 = low, 1 = high), years of age, height (m), sex (0=female, male = 1), L1 English (0 = No, 1 = Yes) and room recording order (0 = Room 1 first, 1 = Room 2 first) as independent variables (Table 1). Each model included a participant random intercept to account for clustering of repeated measures. We report the $\beta$ coefficients for reverberation, labelled *adjusted mean difference*, and their corresponding 95% CI. The regression coefficients represented the standard deviations difference in each speech feature between low and high reverberation rooms. Bootstrap CIs were estimated using a parametric percentile bootstrap with 500 iterations, implemented using the *confint.merMod* method from the *lme4* package for R [29].

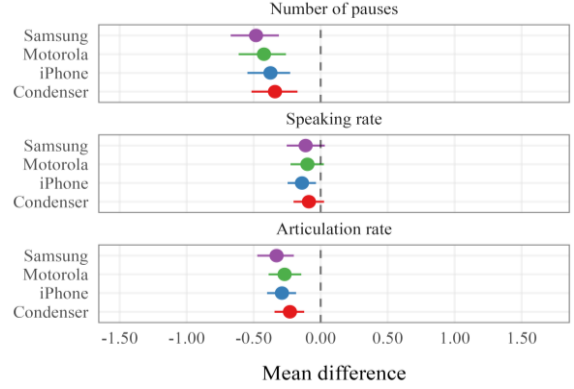

Figure 2: S*tandardized differences in timing features between low and high reverberation per participant over 42 participants. Errors bars are 95% confidence intervals. Negative differences represent lower feature values in the presence of higher reverberation.*

## 3. Results & Discussion

Figures 2 and 3 show the within-person standardized mean differences (and corresponding 95% confidence intervals) in each feature between the two rooms. Positive values indicate that mean feature values were higher in the presence of high reverberation (Room 2) versus low reverberation (Room 1). The three extracted timing features are all affected somewhat by room reverberation (Figure 2). The consistent negative mean difference raises robustness and validity concerns for these features, which are widely used as markers in mental health and neurological speech studies [7].

F0 mean and standard deviation are least affected by reverberation (Figure 3). Both the mean difference and room coefficient have estimates closer to zero and narrow confidence intervals. This is consistent with results reported in [12], [13] where F0 was reported as more robust to ambient noise and microphone type than other speech features. Surprisingly, we observe no clear effect of reverberation on intensity as recorded by the condenser microphone. There are, however, small offsets in intensity in all three smartphones.

Considering F1 and F2, the systematic biases observed in the suprasegmental analysis of the Motorola, iPhone and the condenser microphone are not present in the LME /a/ analysis. We speculate that this is due to the reduction in linguistic variability. Interestingly, the F1 results for Samsung differ from the other devices. We speculate that this could be due to pre-processing on the Samsung. Voice quality features appear the most affected by reverberation, with jitter, shimmer and CPP being clearly affected in all testing scenarios. These results amplify existing concerns about the robustness and validity of voice quality features for health assessment [12], [13]. We observed no clear differences between smartphones, including between the budget phone and other phones we tested. However, we plan to compare devices formally in future work.

In summary, our analysis demonstrates that room reverberation is a source of variability in multiple features. This is unsurprising; reverberation is a convolutive phenomenon and dictated by the raw speech signal and the room impulse response, and almost all extracted features are likely to be somewhat affected. A caveat is that these observations may not be due solely to reverberation; the Lombard effect may have resulted in different raw speech characteristics [30].

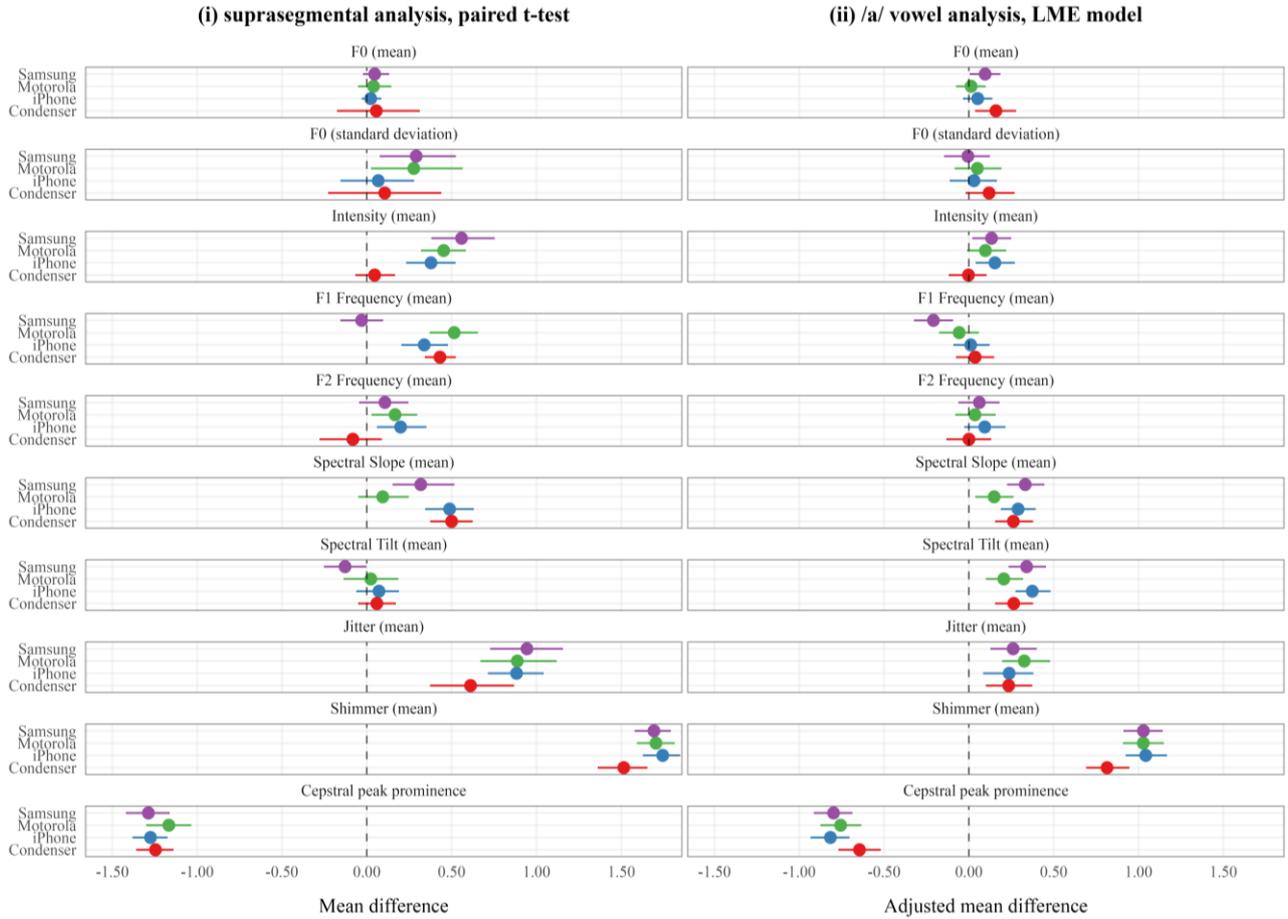

Figure 3: *Standardized differences in features values between low and high reverberation per participant over 42 participants. Mean differences in (ii) are adjusted for participant and measurement variables. Positive differences represent higher values in the presence of higher reverberation. Error bars represent 95% confidence intervals.*

Our study has several limitations. First, it was conducted in healthy volunteers. This was a design choice to begin to investigate reverberation effects while minimizing variability due to pathology and to pilot data collection procedures for future larger studies. Second, our experimental set-up likely represents the upper extreme of reverberation levels in-the-wild and consequently overestimates the effects; however, our results still highlight reliability concerns. Third, we have a limited age range. Fourth, we did not collect sustained vowels as part of our protocol; this might have allowed more reliable extraction of voice quality features. Fifth, we did not compare the effect of different processing tools and algorithm settings. Sixth, our analysis is not a formal comparison of mobile devices. We are planning a larger study to address these.

## 4. Conclusion

Our findings on the effects of reverberation complement observations in the literature on ambient noise and device choice [12], [13]. The speech processing community cannot ignore these sources of variability and expect to develop robust and reliable models for health assessment; otherwise, differences in recording conditions may be attributed to different health states. Instead, we must address the question of how we best record speech in-the-wild that is both robust to variations in recording parameters but still captures salient indicators of an individual's health state. Signal processing and machine learning solutions must also be explored. This could include impulse response estimation techniques and adversarial learning approaches [31, 32]. These are vital steps in the development of transparent, reliable, valid and reproducible tools for health state assessment.

## 5. Acknowledgements

We thank our participants for their support and the KCL Department of Psychology for the use of their test rooms. This work was supported by the MRC Impact Acceleration Account (IAA) King's College London 2021 (MR/X502923/1) and the EPSRC IAA King's College London 2022 (EP/X525571/1). This paper also represents independent research part funded by the National Institute for Health Research (NIHR) Maudsley Biomedical Research Centre at South London and Maudsley NHS Foundation Trust and King's College London and supported by the National Institute for Health and Care Research University College London Hospitals Biomedical Research Centre. Views expressed are those of the authors and not necessarily those of the NHS, the NIHR or the Dept. of Health and Social Care. For T.F. Quatieri: Material is approved for public release, distribution is unlimited, and is based upon work supported by the Under Secretary of Defense for Research and Engineering under Air Force Contract No. FA8702-15-D-0001. Any opinions, findings, conclusions or recommendations expressed in this material are those of the authors and do not necessarily reflect the views of the Under Secretary of Defense for Research and Engineering.